# Thermally Activated Deviations from Quantum Hall Plateaus

M. M. Fogler and B. I. Shklovskii

*Theoretical Physics Institute, University of Minnesota, 116 Church St. Southeast, Minneapolis, Minnesota 55455*

(September 20, 1994 11:36am)

The Hall conductivity $\sigma_{xy}$ of a two-dimensional electron system is quantized in units of $e^2/h$ when the Fermi level is located in the mobility gap between two Landau levels. We consider the deviation of $\sigma_{xy}$ from a quantized value caused by the thermal activation of electrons to the extended states for the case of a long range random potential. This deviation is of the form $\sigma_{xy}^* \exp(-\Delta/T)$. The prefactor $\sigma_{xy}^*$ is equal to $e^2/h$ at temperatures above a characteristic temperature $T_2$. With the temperature decreasing below $T_2$, $\sigma_{xy}^*$ decays according to a power law: $\sigma_{xy}^* = \frac{e^2}{h}(T/T_2)^\gamma$. Similar results are valid for a fractional Hall plateau near filling factor $p/q$ if $e$ is replaced by the fractional charge $e/q$.

The characteristic feature of the quantum Hall effect is that the dependence $\sigma_{xy}(B)$ has a series of plateaus. At $T = 0$ the magnitude of $\sigma_{xy}$ on each plateau is $ie^2/h$, where $i$ is an integer or a simple fraction. The precision of this quantization was verified experimentally to be extremely high (better than one part in $10^9$ [1]). The quantum Hall plateaus serve nowadays as the resistance standard and the most accurate method to evaluate the fine structure constant.

At finite temperatures the deviation $\delta\sigma_{xy} = \sigma_{xy} - ie^2/h$ from the precise quantization appears. This deviation is the subject of this Letter. First of all, we note that directly at the center of the plateau $\delta\sigma_{xy}$ must vanish due to electron-hole symmetry. Therefore, we consider the case when, say, the *electron* contribution to the $\sigma_{xy}$ dominates. This is the case when the Fermi energy is in the gap between two Landau levels but is closer to the upper one by at least several $T$. Hence, our results have implications for metrology mesurements, which are performed in a finite range around the plateau center.

Another interesting problem here is the relation between $\delta\sigma_{xy}$ and $\sigma_{xx}$, which both are very close to zero on the plateau. There have been several theoretical attempts to relate them using quite general arguments and yielding different answers. Amongst them there are the linear relation $\delta\sigma_{xy} \propto \sigma_{xx}$ [2], the quadratic dependence $\delta\sigma_{xy} \propto \sigma_{xx}^2$ [3], and more complicated forms [4]. Experimentally, the relation close to a linear one was observed [1,5].

In this Letter we calculate $\delta\sigma_{xy}$ explicitly for the case where the Landau levels are modulated by a smooth disorder potential $\epsilon(\mathbf{r})$. For definiteness we discuss the integer quantum Hall effect, but the generalization to the fractional Hall effect is straightforward.

If the temperature is not too low so that variable range hopping can be neglected, then both $\delta\sigma_{xy}$ and $\sigma_{xx}$ are determined by electrons thermally activated to the extended state at the classical percolation level $\epsilon = 0$ at the upper Landau level. One would expect

$$\delta\sigma_{xy} = \sigma_{xy}^* \exp(-\Delta/T), \;\; \sigma_{xx} = \sigma_{xx}^* \exp(-\Delta/T), \quad (1)$$

where the activation energy $\Delta$ is the separation between the Fermi energy and the percolation level. The calculation of the prefactor $\sigma_{xy}^*$ is the goal of this paper. We will formulate the result immediately after introducing the model and necessary notations.

Let the correlation length $d$ of the disorder potential be much larger than the magnetic length $\lambda$. One can think of $d$ as the setback distance of the nearby doping layer with charged impurities in a high-mobility heterostructure. The concentration of the activated electrons $n = (1/2\pi\lambda^2)\exp[-(\Delta + \epsilon)/T]$ is exponentially small and we neglect the interaction between them. In the absence of inelastic scattering, guiding centers of electron orbits drift along the contours of constant $\epsilon$, all of which for the exception of the one at $\epsilon = 0$ are closed loops. In this semiclassical picture the tunneling between two loops at the same energy $\epsilon$ is neglected. The tunneling occurs in the vicinity of saddle points, where the loops come close to each other. However, the probability of such tunneling falls off with increasing $\epsilon$ as $\exp(-|\epsilon|/T_1)$, where $T_1 \sim W(\lambda/d)^2$ [6] with $W$ being the standard deviation of $\epsilon(\mathbf{r})$. Below we consider the wide temperature range $T_1 \ll T \ll W$, in which the tunneling can be neglected. We also assume that the inelastic processes are due to the electron-phonon interaction only. The corresponding scattering rate can be readily evaluated using Fermi's golden rule to be $\hbar/\tau_{ph} = \alpha \max\{T, T_{ph}\}$. Here $\alpha = \Xi^2/\rho h s^3 \lambda^2$, is the electron-phonon coupling constant, and $\Xi$ and $\rho$ are the deformation potential constant and the crystal mass density, respectively. For GaAs $\alpha \simeq 0.03(100\text{Å}/\lambda)^2$ [7] and this is the small parameter in this problem. The quantity $T_{ph} = W\lambda/d$ is the characteristic phonon energy. This is because typical hops occur between two contours a distance $\lambda$ from each other. Within this model we obtained

$$\sigma_{xy}^* = \frac{e^2}{h} \begin{cases} c_1(T/T_2)^{8/3}(T_{\mathrm{ph}}/T_2)^{8/7}, & T_1 \ll T \ll T_{\mathrm{ph}} \\ c_2(T/T_2)^{80/21}, & T_{\mathrm{ph}} \ll T \ll T_2 \\ 1, & T \gg T_2, \end{cases}$$
(2)

where $T_2$ is defined as $\alpha^{3/10}W$. For $\sigma_{xx}^*$ we got

$$\sigma_{xx}^* = \frac{e^2}{h} \begin{cases} (1+\alpha^2)^{1/2}, & T_1 \ll T \ll T_2 \\ c_3(T_2/T)^{10/13} + O(\alpha^2), & T \gg T_2. \end{cases}$$
(3)

Here $c_i$'s are numerical factors of order unity. Apart from the terms $\alpha^2$ Eq. (3) was derived within the same model in Ref. [8]. We see that the relation between $\delta\sigma_{xy}$ and $\sigma_{xx}$ is more complicated than a power law. However, the main $T$-dependence comes from the $\exp(-\Delta/T)$ factors and therefore, one can say that the temperature driven dependence $\delta\sigma_{xy}(\sigma_{xx})$ is nearly linear.

We now turn to explanations of our results. Regarding the inelastic scattering we have to distinguish between two regimes. The first one is the diffusive regime for which we can introduce the diffusion coefficient $D \sim \lambda^2/\tau_{\mathrm{ph}}$. This can be done when the energy change $T_{\mathrm{ph}}$ resulting from a typical collision is less than $T$. The second regime is that of single scattering events. It is realized when $T < T_{\mathrm{ph}}$ and will be discussed later.

Consider the diffusive regime first. Introduce the local conductivity tensor with the diagonal components $\sigma_{xx} = e^2 n D/T = \frac{e^2}{h}\alpha \exp[-(\Delta+\epsilon)/T]$ and the off-diagonal components $\sigma_{xy} = -\sigma_{yx} = \frac{e^2}{h}\exp[-(\Delta+\epsilon)/T]$. A non-zero electric field $\mathbf{E}$ produces the spatial variations of the electrochemical potential $\mu(\mathbf{r})$. The current in the system is found from the system of two equations:

$$\mathbf{j} = \frac{1}{e}\overleftrightarrow{\sigma}\boldsymbol{\nabla}\mu, \quad \mathbf{j} = [\hat{z} \times \boldsymbol{\nabla}\psi].$$
(4)

The last equation takes care of current conservation by introducing the stream function $\psi(\mathbf{r})$. With the appropriate boundary conditions and the condition that the spatial average of $\boldsymbol{\nabla}\mu$ is equal to that of $e\mathbf{E}$, this system fully determines $\mathbf{j}(\mathbf{r})$. For this system there exists a remarkable exact relation [9]

$$(\sigma_{xx}^*)^2 + (\sigma_{xy}^*)^2 = (e^2/h)^2(1+\alpha^2),$$
(5)

which holds for a random potential statistically equally distributed around zero. In reality, however, only the symmetrical distribution in the energy band $T$ around the percolation level is needed. Despite the fact that the relation given by Eq. (5) is very useful, it does not shed light on the individual behavior of $\sigma_{xx}^*$ and $\sigma_{xy}^*$. We study this behavior below.

We begin by examining a periodic chess-board geometry where $\epsilon = W \sin\left[\frac{\pi(x+y)}{\sqrt{2}d}\right]\left[\frac{\pi(x-y)}{\sqrt{2}d}\right]$ with the average electric field in the $\hat{y}$-direction (Fig. 1). The regions $\epsilon(\mathbf{r}) \gtrsim T$ (hills) have very low conductance, and, hence, the current avoids them. It flows mainly in the valleys where $\epsilon(\mathbf{r}) < 0$ and crosses from one valley to another via the saddle points (where $\epsilon = 0$).

Once the current passes the saddle point it flows along the slope of the valley. Correspondingly, it deviates from the contour $\epsilon = 0$ towards the more conductive bottom. The current distribution can be inferred from studying two basic elements of the chess-board: a saddle point and a slope. Consider an isolated saddle point $\epsilon(\mathbf{r}) = W(\pi^2/2d^2)(x^2 - y^2)$ first. We choose the boundary conditions $\mu(x, -\infty) = 0$, $\mu(x, \infty) = eU$ and look for the solution that has the full symmetry of the problem: $\mu(x, y) + \mu(-x, -y) = eU$. The desired solution is [10]

$$\psi = \frac{GU}{2}\mathrm{erfc}\left(\frac{x + \beta y}{\sqrt{2}\delta}\right), \quad \mu = \frac{U}{2}\mathrm{erfc}\left(\frac{x - \beta y}{\sqrt{2}\delta}\right),$$
(6)

where $\beta = \alpha + \sqrt{1+\alpha^2}$, $\delta^2 = 2(d/\pi)^2\alpha\beta T/W$. The quantity $G = \frac{e^2}{h}\sqrt{1+\alpha^2}\exp(-\Delta/T)$ can be identified as the conductance of the saddle point.

Our result for $G$ differs only by a small correction $\sim \alpha^2$ from the conductance of a saddle point in the ballistic regime [11,8]. This regime implies that the saddle point connects two "reservoirs" with $\mu = 0$ and $\mu = eU$. At the location of the saddle point, there is a discontinuity in the electrochemical potential. The current passes the saddle point due to a finite $\sigma_{xy}$, and therefore, the conductance of the saddle point is equal to its Hall conductivity. In our case $\mu$ is nearly constant at distances larger than $\delta$ from the saddle point. Thus, if $\alpha$ is small, then the picture of the reservoirs connected by a ballistic saddle point contact is a very good approximation. One observes from Eq. (6) that the current flows through the saddle point in a thin stream of width $\delta$ (Fig. 1). The stream deviates by a small angle $\alpha/2$ from the contour $\epsilon = 0$ towards the bottom of the valley.

Let us resume our study of the chess-board. With $\mathbf{E}$ in the $\hat{y}$-direction, the current through saddle point A (Fig. 1) determines $\sigma_{xx}^*$, and the current through saddle point B gives $\sigma_{xy}^*$. Our solution for an isolated saddle point is of use if the deviation of the current stream on traveling the distance $AB = d$ exceeds the width of the stream $\delta$. In this case only an exponentially small fraction of the current, which enters through saddle point A, leaves the valley through saddle point B.

The current through saddle point A is $GU$, where $U = eEd\sqrt{2}$ is the voltage drop across this saddle point. Similarly, the current through B is almost zero since there is no voltage drop across this saddle point. Therefore,

$$\sigma_{xy}^* = 0, \quad \sigma_{xx}^* = \frac{e^2}{h}\sqrt{1+\alpha^2}.$$
(7)

The equivalent circuit of the chess-board is a square network of identical conductances $\frac{e^2}{h}\sqrt{1+\alpha^2}\exp(-\Delta/T)$ (saddle points) connecting the reservoirs (valleys).

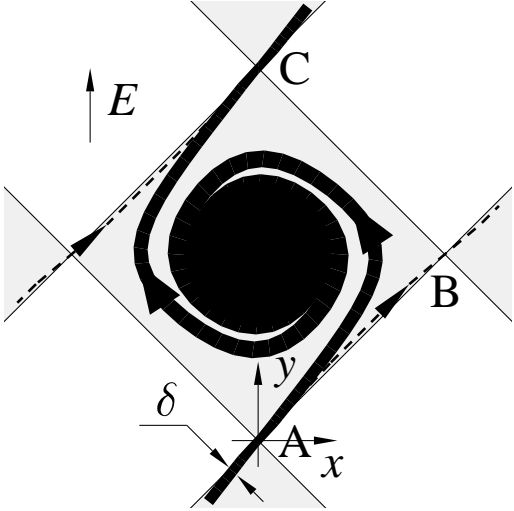

FIG. 1. The picture illustrating our analytical solution of Eq. (4) for a chess-board of valleys (dark) and hills (blank). The currents through saddle points A and B determine $\sigma_{xx}^*$ and $\sigma_{xy}^*$, respectively. The latter current (shown by the dashed line with arrow) is exponentially small. The main current (the black arrow) bypasses B. It enters the valley through saddle points A, then deviates from the perimeter of the valley spiraling counterclockwise towards the interior. It then spirals back out clockwise to exit the valley via saddle point C. The vicinity of saddle point A is described by Eq. (6): the current passes the saddle point in a thin stream of width $\delta$ at a small angle $\alpha/2$ with respect to the valley boundary AB.

To examine the region of validity of Eq. (7), we need to know how a thin stream of current propagates along the second basic element of the chess-board, the slope. For simplicity consider a uniform slope $\epsilon(x, y) = -Wx/d$. The Eqs. (4) may be written as $-D\nabla^2\psi + \mathbf{v}\nabla\psi = 0$, where $\mathbf{v} = \frac{\lambda^2}{\hbar}(W/d)(\alpha\dot{x} + \hat{y})$. This equation is identical to the equation that describes the diffusion of a tracer in a hydrodynamic flow of constant velocity $\mathbf{v}$ [12]. This analogy suggests that the current flows in a thin stream; the stream is directed along $\mathbf{v}$, thus making only a small angle with respect to the contour of constant $\epsilon$; as the current flows away, the stream spreads according to the diffusion law $\delta \sim \sqrt{2Dt}$, where $t = p/u$ is the "time" to travel the distance $p$ from the origin of the stream to a given point. Correspondingly, if the stream originates at $(0, 0)$ then the fraction of current that reaches the point $(0, p)$ is of the order of $\exp[-(v_x t)^2/2\delta^2] = \exp(-p/l)$, where $l = 4dT/\alpha W$. We see that this fraction is exponentially small if $p$ exceeds the "relaxation length" $l$. Applying this consideration to the slope between saddle points A and B (Fig. 1), we find that Eq. (7) is valid if the temperature is low enough: $T \ll T_2^{\mathrm{ch}}$, where $T_2^{\mathrm{ch}} = \alpha W$. In fact, we understand now that in this regime *non-zero* $\sigma_{xy}^*$ *appears only due to the fraction of current propagating along the path* AB. Our calculation gives $\sigma_{xy}^* = \frac{e^2}{h}\sqrt{8T/\pi T_2^{\mathrm{ch}}}\exp(-T_2^{\mathrm{ch}}/2T)$. Similarly, $\sigma_{xx}^*$ differs from the value given by Eq. (7) due to the current propagating along the twice longer path ABC (Fig. 1). Correspondingly, the correction to $\sigma_{xx}^*$ is of the order of $(\sigma_{xy}^*)^2$ (both in units of $e^2/h$) in agreement with Eq. (5).

Finally, we examine a random system. We are going to show that in this case $\sigma_{xy}^*$ exhibits not an exponential but only a power law smallness. First we argue that the reservoirs in a random system will be not single valleys but collections of many of them with the total spatial extent much larger than $d$. Indeed, the random system may be viewed as a chess-board whose saddle point heights $\epsilon_i$'s are randomly distributed around zero with the standard deviation of order $W$. Now the conductances of the equivalent network will have a very broad range of values $\frac{e^2}{h}\exp[-(\Delta + \epsilon_i)/T]$. The percolation approach suggests that the current in such a network is carried by a critical subnetwork consisted of saddle points with heights $-W < \epsilon_i \lesssim T$ [13]. This subnetwork is topologically equivalent to an infinite cluster whose sites are associated with the valley bottoms and whose bonds come through the saddle points specified above. The backbone of the infinite cluster consists of nodes separated by a typical correlation length $\xi = d(W/T)^\nu$, where $\nu = 4/3$ [13,12]. The nodes are connected by essentially one-dimensional paths made of both singly connected bonds, called "red bonds", and multiply connected parts, called "blobs". Amongst the red bonds the most important are those that come through saddle points with $0 < \epsilon_i < T$. We call them critical bonds. There is typically *only one* critical bond per path [13]. Consider now a collection of bonds, which depart from some node and then follow the paths outgoing from this node all the way until critical bonds are met. This set of bonds defines a reservoir in the random system. Indeed, by construction the critical bonds contain the largest unshunted resistors; therefore, the drop of the electrochemical potential is large only across the critical bonds. The saddle points of the critical bonds are, hence, the contacts that connect a given reservoir to other ones.

The typical diameter of the reservoir is $\xi$ and its perimeter (or hull) is $p = d(\xi/d)^{d_{\mathrm{H}}}$, where $d_{\mathrm{H}} = 7/4$ is the fractal dimension of the hull [12]. In the spirit of our analysis of the chess-board, we can say that the fraction $\exp(-p/l)$ of the current entered through a saddle point A can exit through the neighboring saddle point B (Fig. 2). This fraction is exponentially small if $T \ll T_2$, $T_2 = \alpha^{3/10}W$. It means that, as in the case of a chess-board, the equivalent circuit is the network of independent conductances and Eq. (7) applies. *But the conclusion that $\sigma_{xy}^*$ is also exponentially small would be incorrect.* The reason is that in a random system the distance $p$ along the perimeter between two critical saddle points might be as small as $l$, i.e., much less than a typical one. Such an event gives rise to a "Hall generator" (Fig. 2) contributing $\sim e^2/h$ to $\sigma_{xy}^*$. The Hall conductivity of the sample will then be proportional to the probability of finding such a rare generator in a square with the sides of length $\xi$ [14]. Our estimate of this probability is as follows. First we convert the distance along the perimeter into the spatial separation: our critical saddle points must be within the distance $l_0 = d(l/d)^{1/d_H}$ from each other. Choose a critical saddle point in a square with the side $\xi$. Under the assumption that the positions of the critical saddle points are uncorrelated, the probability that another critical saddle point is situated within a circle of radius $l_0$ around a given one is $(l_0/\xi)^2$ since the density of the critical saddle points is $1/\xi^2$. This estimate leads to the result $\sigma_{xy}^* = \frac{e^2}{h}(T/T_2)^{80/21}$ [Eq. (2)].

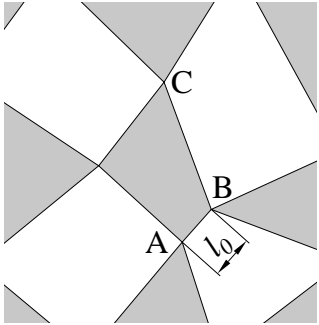

FIG. 2. The random system may be viewed as a distorted chess-board of Fig. 1. A finite probability for the side AB of a reservoir to be as small as $l_0$ gives rises to a "Hall generator" contributing to $\sigma_{xy}^*$.

Consider now the case $T_1 \ll T \ll T_{ph}$. The diffusion picture where electrons can hop many times up and down the slope is no longer valid. Instead, the electrons hop predominantly downward. The relaxation of the current along the perimeter is described now by the probability of avoiding scattering by phonons: $\exp(-t/\tau_{ph})$. Correspondingly, the relaxation length $l = v\tau_{ph} = \lambda^2 W/\alpha d$ is temperature independent. This explains the change in the exponent from $80/21$ to $8/3$ in Eq. (2).

Finally, we briefly discuss the transport at temperatures away from the range $T_1 \ll T \ll T_2$. If $T \lesssim T_1$ tunneling and variable range hopping become important. This will be the topic of a separate paper. In the opposite limit $T \gg T_2$, $\sigma_{xy}$ approaches $e^2/h$. Indeed, at these temperatures the deviation of the current stream from the contour $\epsilon = 0$ is much smaller than the width of the stream: $\alpha p \ll \delta$. Correspondingly, the effect of the current stream sliding down to the reservoirs is small. The current flows mainly within the strip of width $\delta$ centered around the percolation trajectory $\epsilon = 0$. It may be verified that within this region $|\epsilon| \ll T$ and, therefore, the Hall conductivity $\sigma_{xy} = \frac{e^2}{h}\exp[-(\Delta + \epsilon)/T]$ is approximately constant. In this case the sample averaging of $\sigma_{xy}$ necessary for calculating of $\sigma_{xy}^*$ trivially gives $\sigma_{xy}^* = e^2/h$.

In conclusion we note that on on a fractional Hall effect plateau at filling factor slightly exceeding $p/q$ one can apply our theory to the transport of quasielectrons of fractional charge $e/q$. This leads only to the substitution $e$ by $e/q$ in Eqs. (2,3). Unfortunately, in high mobility samples the temperature dependence of only $\sigma_{xx}$ and only at its minima was studied (see the bibliography of Ref. [8]). We propose to investigate the temperature dependence of *both* $\delta\sigma_{xy}$ and $\sigma_{xx}$ *away from* the very minima of $\sigma_{xx}$ to verify the predictions of Eqs. (2,3) and the relation of Eq. (5). On the other hand, this kind of data is available for low mobility samples [1,5] where the random potential is supposedly short range. In such systems at high enough temperatures (before the onset of hopping conductivity) both $\sigma_{xx}$ and $\delta\sigma_{xy}$ are determined by the band of delocalized states around $\epsilon = 0$. While the former was calculated in Ref. [15], the Hall conductivity $\delta\sigma_{xy}$ for this case will be discussed elsewhere.

Useful discussions with I. M. Ruzin and D. G. Polyakov are greatly appreciated. This work was supported by NSF under Grant No. DMR-9321417.